\documentclass{aa}
\usepackage{psfig}

\def\ros{{\sl ROSAT }}

\def\G{$\Gamma_{\rm x}$ }
\def\NH{$N_{\rm H}$ }

\def\degs{\ifmmode ^{\circ}\else$^{\circ}$\fi}
\def\amin{\ifmmode ^{\prime}\else$^{\prime}$\fi}
\def\asec{\ifmmode ^{\prime\prime}\else$^{\prime\prime}$\fi}
\def\farcs{\hbox{$.\!\!^{\prime\prime}$}}  

\def\approxlt{\mathrel{\hbox{\rlap{\lower.55ex \hbox {$\sim$}}
        \kern-.3em \raise.4ex \hbox{$<$}}}}
\def\approxgt{\mathrel{\hbox{\rlap{\lower.55ex \hbox {$\sim$}}
        \kern-.3em \raise.4ex \hbox{$>$}}}}
\begin{document}
 
   \thesaurus{03         
              (11.01.2;  
               11.09.1;  
               11.17.2;  
               11.19.3;  
               13.25.2)  
}
   \title{The giant X-ray outbursts in NGC\,5905 and IC\,3599: \\
          Follow-up observations and outburst scenarios}
   \author{Stefanie Komossa\inst{1}, Norbert Bade\inst{3,2}} 

   \offprints{S. Komossa, skomossa@mpe.mpg.de}
 
  \institute{Max-Planck-Institut f\"ur extraterrestrische Physik,
             85740 Garching, Germany
  \and 
             Hamburger Sternwarte, Gojenbergsweg 112, 21029 Hamburg, Germany 
  \and   
             Schwalbenweg 55, 89081 Ulm, Germany 
       }

   \date{Received 28. July 1998; accepted 4. December 1998}
   \maketitle\markboth{St. Komossa, N. Bade~~~ The giant X-ray outbursts in NGC\,5905 and IC\,3599}{}  

   \begin{abstract}
Huge amplitude X-ray outbursts in a few galaxies 
were reported in the last few years. 
As one of the exciting possibilities to explain these observations,
tidal disruption of a star by a central supermassive black hole
has been proposed. 
In the present paper, we perform a detailed discussion of this and other 
possible 
scenarios for the X-ray outburst in NGC\,5905,
 and a comparison of NGC\,5905 and IC\,3599 in outburst as well as in quiescence.
To this end we present 
(i) a thorough analysis of all \ros  
PSPC X-ray observations of NGC\,5905 and
new HRI data,
(ii) optical photometry of NGC\,5905 quasi-{\em simultaneous}
to the X-ray outburst and on longer terms, 
(iii) the first post-outburst optical spectra of NGC\,5905 and    
high-resolution post-outburst spectra of IC\,3599,  and 
(iv) photoionization models for the high-excitation emission lines
that were discovered in the optical outburst spectrum of IC\,3599.
The investigated outburst models include, besides the tidal disruption
event,  a supernova in dense medium, 
an accretion-disk instability, 
an event of extreme gravitational lensing,   
the X-ray afterglow of a GRB, 
and the possibility of a warm-absorbed hidden Seyfert nucleus in the
center of the galaxy. The successful models, 
all involving the presence of a central supermassive black hole, 
are selected and implications are discussed. 
The optical spectra of both galaxies in quiescence 
are carefully examined for signs of {\em permanent} low-level 
Seyfert-activity. Whereas IC\,3599 shows several signs of activity,
none is revealed for NGC\,5905. At present, and among the X-ray bursts,
this makes NGC\,5905 the only
safe candidate for a tidal disruption event in an otherwise {\em non-active}
galaxy.   
The physical conditions in the HII emission-line gas   
are also investigated. 
We briefly comment on a search for further highly variable
objects on the basis of \ros observations. Several with factors 10-20
are found; all of them are known to harbour warm absorbers. 
   

   \end{abstract}
 
\section{Introduction}

X-ray variability by factors of $\sim$2 -- 3 is a common property of active
galaxies (e.g., Mushotzky et al. 1993). However, X-ray outbursts by factors of
order 100 are extremely rare; only very few objects have been 
reported to show such behaviour. Among these are IC\,3599 (Brandt et al. 1995,
BPF95 hereafter; Grupe et al. 1995a, G\&95 hereafter) and NGC\,5905 (Bade et al. 1996,
paper\,I hereafter).  

Both galaxies were detected in the X-ray high-state during the \ros all-sky survey
and then declined in intensity within months--years. The peak luminosities 
exceeded $L_{\rm x}$ = 10$^{42}$ erg/s and the outburst spectra were very
soft (photon indices \G $\approx$ --4). Further, both objects appear
rather inactive, as judged from optical spectra (this point will be examined
in detail below). 
Here we focus on NGC\,5905, and perform a comparison with IC\,3599.  
For both sources, tidal disruption of a star by a supermassive
black hole (SMBH) has been mentioned
as possible outburst mechanism (BPF95, G\&95, paper I). 
The observation of a UV flare in the center of the elliptical galaxy NGC\,4552
was interpreted in the same sense (Renzini et al. 1995). 

The possibility of tidal disruption of a star by a SMBH was originally
proposed as a means of fuelling active galaxies
(Hills 1975), but was later dismissed. 
Peterson \& Ferland (1986) suggested this mechanism as possible explanation for
the transient brightening and large width of the HeII line observed in a Seyfert galaxy. 
Tidal disruption was invoked by Eracleous et al. (1995) in a model 
to explain the UV properties of LINERs.
Rees (1988, 1990)
suggested that the flare of electromagnetic radiation predicted
when the stellar debris is swallowed by the black hole could be
used as a tracer of SMBHs in nearby {\em non-active} galaxies.
This question is of particular interest in the context of AGN evolution; i.e.,
what fraction of galaxies have passed through an active phase, 
and how many now have non-accreting and hence unseen SMBHs at their centers
(e.g., Rees 1989)? 
Several approaches were followed to study this question.
Much effort has 
been put in deriving central object masses from studies of the dynamics of 
stars and gas in the nuclei of nearby galaxies.                 
Earlier groundbased evidence for central quiescent dark masses
in non-active galaxies  
(e.g., Tonry 1987,     
Dressler \& Richstone 1988,     
Kormendy \& Richstone 1992)     
has been strengthened
by recent HST results (e.g., van der Marel et al. 1997,     
Kormendy et al. 1996;   
see Kormendy \& Richstone 1995 for
a review). Evidence for a SMBH in our galactic center
is strengthening as well (Eckart \& Genzel 1996).      

{\em X-rays} trace the very vicinity of the SMBH and provide
an important probe of the AGN's central regions.
Information can be extracted from  
X-ray variability and continuum shape (e.g., Mushotzky et al. 1993)
and strengths and profiles of X-ray emission lines 
(e.g., Tanaka et al. 1995, Bromley et al. 1998).
This X-ray approach was hitherto mainly applied to galaxies that were previously known
to be active.{\footnote{One 
exception, the presumed HII galaxy IRAS\,15564+6359 that showed  
X-ray variability (Boller et al. 1994)
turned out to be active, as judged by  
optical spectra (Yegiazarian \& Khachikian 1988, Halpern et al. 1995).}}

Presenting new observations and outburst models 
in the present study we critically assess the possibility
of an event of tidal disruption in NGC\,5905 (and IC\,3599) 
and contrast this with other possible outburst
scenarios, some (but not all) of which involve the presence of a SMBH as well.  

The paper is organized as follows: In Sect. 2 we describe the
data reduction. Section 3 deals with the X-ray spectral analysis
and in Sect. 4 we provide the optical observations of NGC\,5905 and IC\,3599. 
Section 5 presents photoionization models for the optical outburst 
emission-line spectrum of IC\,3599. 
Sections 6--9 are concerned with the discussion of the data. In Sect. 6
we scrutinize scenarios for the X-ray outburst in NGC\,5905, Sect. 7 
investigates the properties of NGC\,5905 in quiescence, in Sect. 8 
we perform a comparison with IC\,3599, and in Sect. 9 we briefly
present results from a search for further X-ray outbursters.    
Section 10 provides a summary and the conclusions. 

Luminosities are given for $H_{\rm o}$=50 km/s/Mpc
if not stated otherwise.  

\section{Data reduction}

\subsection{X-ray data}

\paragraph {PSPC.}
The \ros PSPC (Tr\"umper 1983, 1990; Pfeffermann et al. 1987)
data reduction for NGC\,5905 was described in paper I. In the same
manner, we treated the PSPC all-sky survey data of IC\,3599. In brief, photons
were extracted in a circular cell around the target source, the background 
was determined from a source free region along the scanning direction of
the telescope and corrected for.    
The data reduction was performed using the EXSAS software package
(Zimmermann et al. 1994a).

\paragraph {HRI.} 
An HRI observation of NGC\,5905 was performed from November 18--21, 1996, 
centered on the target source.   
The total exposure time is about 76 ksec. X-ray emission from NGC\,5905 is detected
with a background-corrected countrate of 0.0007$\pm{0.0001}$ cts/s. 
This indicates a further drop in brightness by a factor $\sim$2 as
compared to the 3-yr earlier last PSPC observation.  
The longterm X-ray lightcurve
is displayed in Fig. \ref{outlight}. There, we converted the HRI countrate 
to PSPC countrate assuming constant spectral shape.

\subsection{Optical spectra}

In March 1997 we took spectra of NGC\,5905 and IC\,3599 with
the focal reducer MOSCA at the Calar Alto 3.5m
telescope in the green ($\rm 4150 - 6650\AA$) and red ($\rm 5800 - 8300\AA$)
wavelength region. For NGC\,5905 the slit width was 1\arcsec\ which yields a
spectral resolution of 3\,\AA, in the case of IC\,3599 the slit width was
$1\farcs 5$ and the spectral resolution is 4.5\,\AA. In both cases the slit was
oriented in E-W direction.
The optical data were reduced within the MIDAS
package. The wavelength calibration was performed with Hg-Ar-Ne comparison
frames and the velocity zero point was set by the night sky lines on the galaxy
frames. Relative intensity calibration was achieved by taking spectra of the
spectrophotometric standard star GD71.

\section{X-ray spectral analysis} 

\subsection{NGC 5905}

\subsubsection{Standard models}

In addition to the analysis in paper I we add several further models,
which we later use in discussing possible outburst scenarios.
The results for the powerlaw (pl) and pl+blackbody (bb) model fit 
of paper I are summarized in Tab. 1.
In the following, we first fit the high-state data due 
to better photon statistics, and then comment on the low-state spectrum.
If not stated otherwise, \NH was fixed to the Galactic value, 
$N_{\rm H}^{\rm gal} = 1.47~10^{20}$ cm$^{-2}$ (Dickey \& Lockman 1990). 
Applying a single bb, we find a temperature $kT$ = 0.06 keV, an intrinsic
(0.1-2.4 keV) luminosity $L_{\rm x}$ = 2.6 10$^{42}$ erg/s, and 
$\chi^{2}_{\rm red}$ = 1.2. 
The corresponding values for emission from a Raymond-Smith (rs) plasma
are $kT$ = 0.1 keV, $L_{\rm x}$ = 2.2 10$^{42}$ erg/s, and 
$\chi^{2}_{\rm red}$ = 1.2.
If \NH is treated as free parameter we get \NH = 4.4 10$^{20}$ cm$^{-2}$,
$kT$ = 0.06 keV, $L_{\rm x}$ = 44.5 10$^{42}$ erg/s, and
$\chi^{2}_{\rm red}$ = 1.1. 
Fitting the  accretion disk model included in EXSAS yields
for a BH mass of 10$^4$ M$_{\odot}$ 
an accretion rate in terms of the Eddington rate of ${\dot m}$ = 0.2, 
$L_{\rm x}$ = 2.9 10$^{42}$ erg/s, and 
$\chi^{2}_{\rm red}$ = 1.1.
Some models are shown and compared in Fig. 1. 
\begin{table*}[ht]
  \caption{Summary of some X-ray spectral fits to IC\,3599 and NGC\,5905 
           (survey=high-state observation,
           pointing=low-state).}
  \begin{tabular}{lllllllllll}
  \noalign{\smallskip}
  \hline
  \noalign{\smallskip}
 galaxy &  model $\rightarrow$  & \multicolumn{3}{l}{pl, \NH free~} &
                       \multicolumn{2}{l}{pl$^{(1)}$~~~~~~~} &
                         \multicolumn{3}{l}{pl + bb$^{(1)}$} & ref \\
  \noalign{\smallskip}
  \hline
  \noalign{\smallskip}
   &  & $N_{\rm H}^{(2)}$ & \G & $\chi^2_{\rm red}$ & \G & $\chi^2_{\rm red}$ & 
                                       \G & $kT_{\rm bb}^{(3)}$ & $\chi^2_{\rm red}$ &\\ 
  \noalign{\smallskip}
  \hline
  \hline
  \noalign{\smallskip}
  \noalign{\smallskip}
 IC\,3599 & survey & 5.2 & --4.8 & 1.5 & --3.1 & 6.4 & --2.2 & 91 & 2.2 & BPF95\\
   & & & & & --3.1 & 3.1 & --2.0$^{(4)}$ & 91 & 1.4 & G\&95 \\  
  \noalign{\smallskip}
 &   pointing II & 2.9 & --4.6 & 0.9 & & & & & & BPF95 \\ 
  & & & & & --3.3 & 1.2 & --2.0$^{(4)}$ & 65 & 1.0 & G\&95 \\
  \noalign{\smallskip}
  \hline 
  \noalign{\smallskip}
  \noalign{\smallskip}
 NGC\,5905 & survey & 2.9 & --5.1 & 1.0 & --4.0 & 1.26 & --1.9$^{(4)}$ & 48 & 0.9 & paper I\\
  \noalign{\smallskip}
 &  pointing &  8.2 & --4.5 & 1.0 & --2.4 & 1.2 & & & & \\
  \noalign{\smallskip}
  \hline
  \noalign{\smallskip}
     \end{tabular}
  \label{tab2}

  \noindent{\small
 $^{(1)}$$N_{\rm H}$
fixed to $N_{\rm H}^{\rm Gal}$ (= 1.47, 1.3 10$^{20}$cm$^{-2}$ for NGC\,5905 and IC\,3599, respectively),
 ~ $^{(2)}$in 10$^{20}$cm$^{-2}$, $^{(3)}$in eV, $^{(4)}$fixed 
}
\end{table*}

\subsubsection{Warm absorber models}  

The presence of a warm absorber could explain the observed variability
and softness of the X-ray spectrum of NGC\,5905 (Komossa \& Fink 1997a).
This model is discussed further in Sect. 6.5.
To test for the presence of a warm absorber, we have applied our {\em Cloudy}-based
models (described e.g. in Komossa \& Fink 1997b-d; for details on the
photoionization code {\em Cloudy} see Ferland 1993, 1998).
The properties of the ionized material which are derived from X-ray spectral fits are
the total hydrogen column density $N_{\rm w}$ of the
warm absorber and the ionization parameter $U$, defined as
\begin {equation}
U = Q/(4\pi{r}^{2}n_{\rm H}c)~,~~ 
{\rm with}~~  Q    
\propto  \int\limits_{\nu_0}^{\infty} { f_{\nu} \over {h\nu}} \rm{d}\nu
\end {equation} 
where $f_{\nu}$ is the flux of the incident continuum, $\nu_{\rm 0}$ the 
frequency at the Lyman limit.
We used a mean Seyfert spectrum of piecewise powerlaws  
with $\alpha_{\rm uv-x}$=--1.4 in the EUV and \G = --1.9.
Solar abundances (Grevesse \& Anders 1989)
were adopted if not stated otherwise.

The fit of a warm absorber to the high-state spectrum proves to be successful
with an ionization parameter of log $U \simeq$ 0.0 and a column density 
log $N_{\rm w} \simeq$ 22.8 ($\chi^2_{\rm red}$ = 1.0).
In a second step, we applied the model of a {\em dusty} warm absorber. 
The presence of dust could 
obscure the nucleus at optical/UV wavelengths, and explain why the optical spectrum
of NGC\,5905 looks like that of a non-active galaxy (for details see below).

The dusty models were again calculated with {\em Cloudy}.
We first assumed a Galactic gas/dust ratio, the dust species graphite and silicate,
and depleted gas-phase metal abundances. 
The photon index of the intrinsic X-ray spectrum was fixed to the canonical Seyfert value of 
\G=--1.9. No fit is possible. This can be traced back to the strong `flattening
effect' of dust (Komossa \& Fink 1997b-d; Komossa \& Bade 1998) 
that is in conflict with the steep observed spectrum.           
Several models with modified dust properties were explored. 
If we allow only the silicate species to be present, being depleted by a factor of 
$\sim$20, it becomes  possible to fit the high-state spectrum
(with log $U \simeq$ 0.2, log $N_{\rm w} \simeq$ 22.9).

\subsubsection{Combined fits}

Here we fit the low-state data, fixing some of the parameters to 
the values determined for the high-state 
observation. This is to test whether the data can be explained by requiring the
change of one parameter only, or as few as possible. 
The following descriptions were applied: \\
(i) Pure absorption variability. Is the decline in luminosity 
caused by increased absorption,  
e.g., by a cloud blocking the line-of-sight (l.o.s.)
during the low-state or by a transitory 
decrease  
in absorption during the high-state ?   
To test this possibility, the pl index and norm were fixed to the values determined
for the high-state, and the low-state was fit with $N_{\rm H}$ as free parameter.
The quality of the
fit turns out to be unacceptable ($\chi^2_{\rm red}$=2.8). \\
(ii) Changes in warm absorption. Ionized material responds to changes in the impinging
photon flux.
The observed spectrum shows a flattening with increasing time from the outburst.
Whereas a warm absorber is expected to first respond by deeper
absorption (causing a steeper observed spectrum)
to a decline in intrinsic luminosity 
it then goes from `warm' to `luke-warm', shifting the edges to more
lowly ionized species,
thereby flattening the observed spectrum.
In any case, one expects lower $U$ in the low-state. 
Whereas a warm absorber can formally be fit to
the low-state spectrum, we find a {\em higher} ionization parameter (log $U \simeq 0.5$)
as well as higher column density (log $N_{\rm w} \simeq 23.2$).  \\
(iii) Same bump component, declined in intensity and/or temperature.
Single component fits (like bb or rs) already show that an {\em in}crease in temperature
(in addition to the decrease in luminosity) is found for all models, if $T$
is left as free parameter. If constant $T$ is enforced, i.e. fixed
to the value determined for the
high-state observation, $\chi^2$ is unacceptably large ($\chi^2_{\rm red} \approxgt 2$). 
In the accretion disk description, if the black hole mass
is fixed, ${\dot m}$ = 1 in the low-state, and
the fit is still unacceptable. \\
(iv) Variable bump component on top of constant underlying pl.
Fixing the norm of the underlying pl to the  high-state value (in the pl+bb description)
it is not possible to fit the low-state. Such a two-component description of
the low-state anyway does not lead to an improvement of the fit. If it
is nevertheless applied, a successful fit requires the powerlaw component to decline
by about a factor of 10 in flux, and $kT_{\rm bb}$ to decrease to 12 eV
(as compared to 48 eV in the high-state). 

Since none of the `tight' descriptions proved to be successful, we conclude
that the spectral component seen in the PSPC low-state 
(with $L_{\rm x} \simeq 4 \times 10^{40}$ erg/s
and \G $\simeq -2.4$) is not the same as the one
in the high-state,  
and most probably dominated by the emission of 
the host galaxy,
which is typically of the order $10^{38-40}$ erg/s in non-active
galaxies (e.g., Fabbiano 1989, Vogler 1997).

  \begin{figure}[h]
\psfig{file=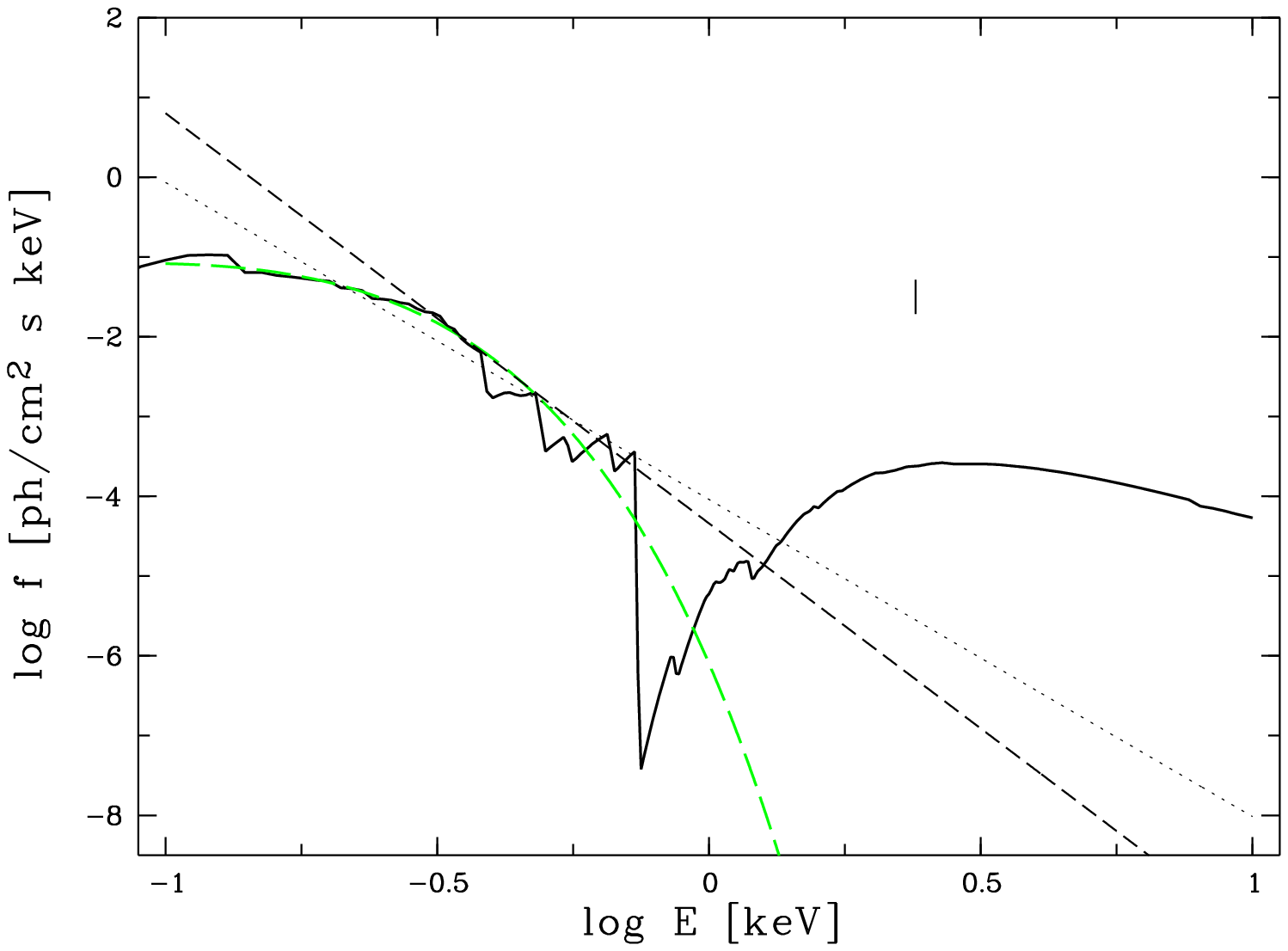,width=8.9cm}
\vspace{0.2cm}
\psfig{file=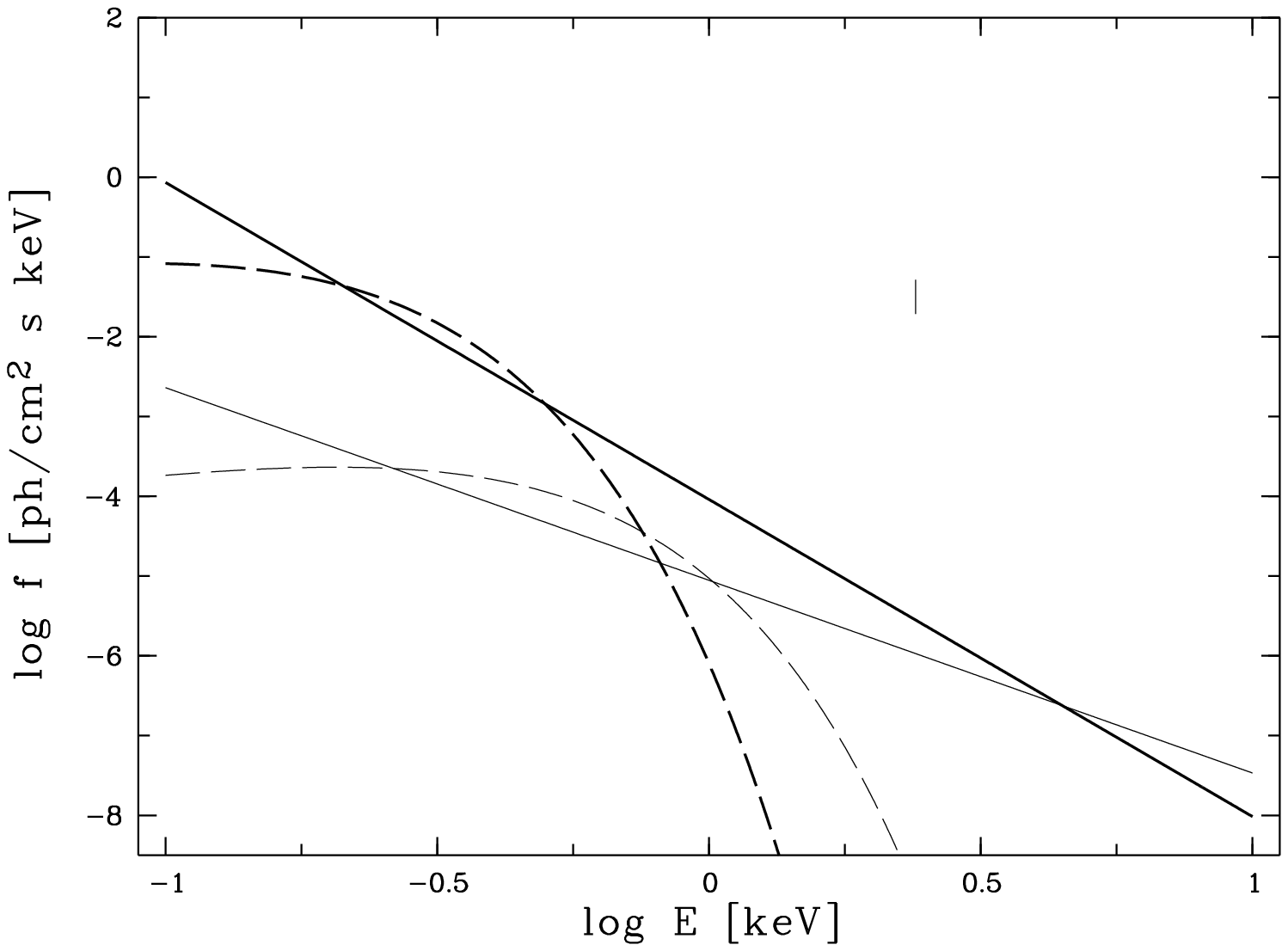,width=8.9cm}
 \caption[allmod]{
Comparison of several selected spectral fits to the high-state and low-state data
of NGC\,5905. 
{\bf Upper panel}: Different descriptions of the high-state spectrum;
solid line: warm absorber, thin long-dashed line: black body, dotted line: powerlaw.
In all these models, \NH was fixed to the Galactic value.  
The short-dashed line corresponds to a powerlaw with \NH treated as free parameter.
Whereas the models are rather similar in the soft energy region 
(e.g., compare the bb with the warm absorber), they show very different high-energy behaviour.
In fact, the single thermal models tend to underpredict the observed flux above 1 keV.
{\bf Lower panel}: Low-state data as thin lines; solid line: powerlaw, dashed line:
black body. For comparison, with thick lines the same models for the high-state 
data are re-plotted. 
}
\label{allmod}
\end{figure}

\subsection{IC\,3599} 

Standard fits (pl and bb) are summarized in Tab. 1. 
Here, we comment on the possible presence of a warm absorber
since no such model has been applied previously.  
Fitting this model to the high-state X-ray spectrum
(with \G=--1.9 fixed) 
yields log $U \simeq 0.6$ , log $N_{\rm w} \simeq 23.3$
and cold absorption consistent with the Galactic value. The fit comes 
with a rather high $\chi^2_{\rm red}$ = 2.4.  
However, none of the applied models gives an acceptable fit (cf. Tab. 1) 
the warm absorber being  
as successful as the pl + bb description. 

\section{Optical observations} 

\subsection {Photometry}

Depending on the ratio $L_{\rm opt}/L_{\rm x}$ of the variable component,
we might expect to see some optical variability as well; a factor 100
in the {\em variable} component would correspond to $\sim$5$^{\rm m}$
(but `smeared' due to seeing).
To search for optical variability, we used the 
photographic plates available at {\sl Sonneberg Observatory}.
For an overview of the program and plate details 
see Br\"auer \& Fuhrmann (1992). 
We examined all plates taken around the time of the observed X-ray
outburst as well as a random sample of plates 
taken on a longer timescale 
(between the years 1962 and 1995). We mainly used the blue plates
due to their better quality; in the time-interval of the X-ray observation
we also checked the red plates.   

The brightness of the nucleus of NGC\,5905 was measured 
relative to the nucleus of the similarly bright nearby non-active
spiral galaxy NGC\,5908.  
%
  \begin{figure*}[ht]
\psfig{file=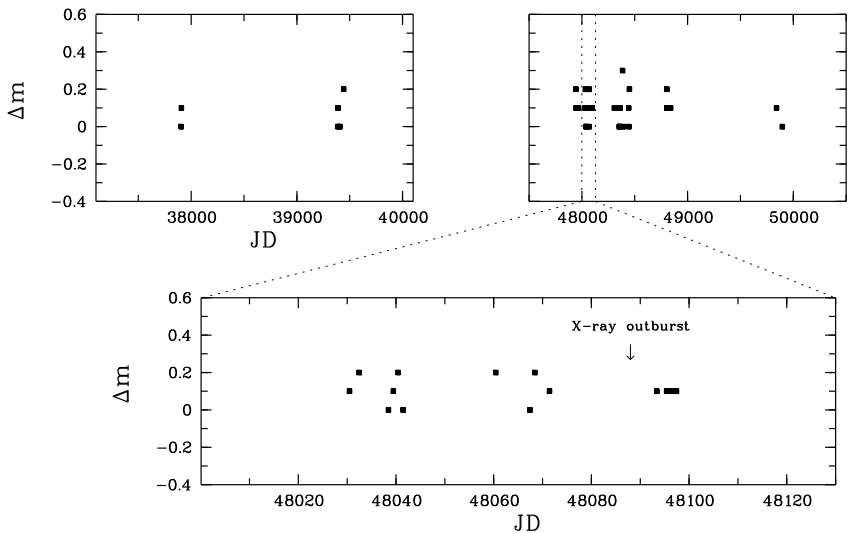,width=11.2cm}      
\hfill
\begin{minipage}[]{0.32\hsize}\vspace*{-7.2cm}   
\hfill
 \caption[n5905_sonne]{
Optical lightcurve of the nucleus of NGC\,5905, based on
photographic plates taken at {\sl Sonneberg Observatory}.
The abscissa gives the Julian date in JD-2400000 (the data points bracket the
time interval 1962 -- 1995).
The upper panel shows the longterm lightcurve, the lower one the lightcurve in the months
around the X-ray outburst.
Within the error of about 0.2$^{\rm m}$
the optical brightness is constant.
}
\label{n5905_sonne}
\end{minipage}
\end{figure*}
%
The lightcurve is shown in Fig. \ref{n5905_sonne}.
Within the error of $\sim$0.2$^{\rm m}$ the optical brightness
is constant. This holds for the time around the X-ray outburst as well as on 
long terms, and places tight constraints on the outburst scenarios (Sect. 6).
 
\subsection {Optical post-outburst spectra}

  \begin{figure*}[thbp]
\psfig{file=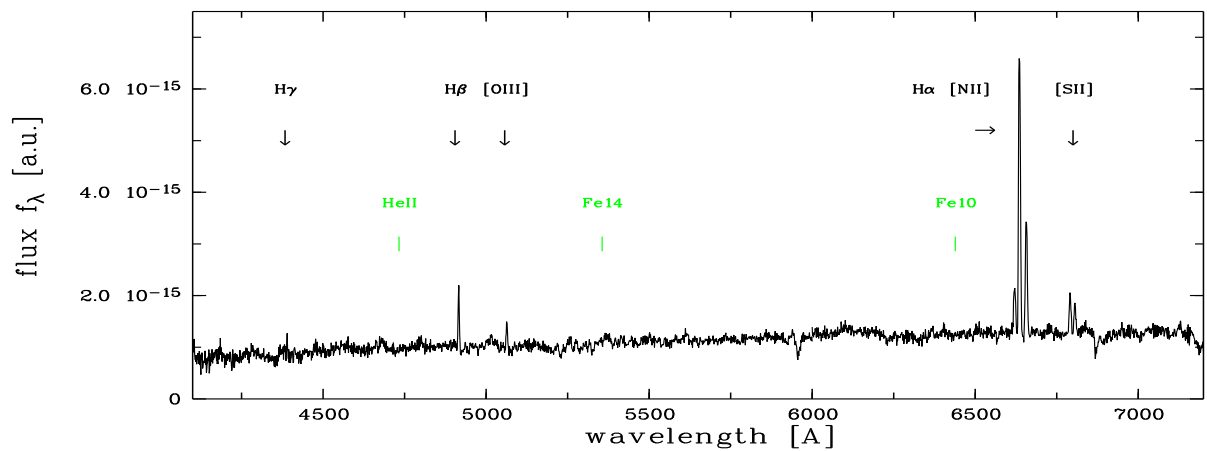,width=16.1cm,height=6cm}
 \caption[n5905_opt]{ 
Optical spectrum of the nucleus of NGC\,5905 taken
about 6.7 years after the X-ray outburst.  
The arrows mark the detected emission lines.
The weakness of [OIII]$\lambda$5007 compared to H$\beta$ 
and the narrow lines classify the galaxy
as HII-type.
The positions of some high-ionization lines, not detected in the spectrum,
are marked as well. 
}
\label{n5905_opt}
\end{figure*}

\subsubsection{NGC\,5905} 

\paragraph{Emission line strengths.}
The optical spectrum of NGC\,5905 is displayed in Fig. \ref{n5905_opt}.
It shows strong emission lines on a continuum with absorption lines
produced by the stellar content of the nucleus. 
The contribution of the stellar content is higher in NGC\,5905
than in IC\,3599 (judged by the deeper absorption in the former's spectrum).
The Balmer lines are composed of an emission line plus a broader
absorption which hints to a large portion of moderately young stars in the
nucleus of NGC\,5905 (stars of spectral type B, A and F). 
Further detected emission lines are those of [OI], [OIII], [NII] and [SII]
(Tab. 2). In Fig. \ref{diagn} emission line ratios are plotted in 
the diagnostic diagrams of Veilleux \& Osterbrock (1987). 
No high-ionization lines like HeII or [FeX] were detected.
 
   \begin{table}[h]
\caption{Optical emission line ratios relative to H$\beta_{\rm qu}$ (index `qu' for `quiescence')
and line widths (for the stronger
lines only) of NGC\,5905 und IC\,3599 in quiescence. (For comparison: BPF95 measure
for the outburst spectrum of IC\,3599 FWHM$_{\rm H\beta} \simeq 1200$ km/s and
FWHM$_{\rm H\alpha} \simeq 1500$ km/s.)
Results given in the Tab. were derived by fitting a one-component Gaussian to each emission line;
see text for details.
}
     \label{obsres}
      \begin{tabular}{rcccc}
      \noalign{\smallskip}
      \noalign{\smallskip}
      \hline
      \noalign{\smallskip}
       Line~~~~  & \multicolumn{2}{c}{$I/I_{\rm H\beta_{\rm qu}}$} & \multicolumn{2}{c}{FWHM [km/s]} \\ 
       \noalign{\smallskip}
      \hline
       \noalign{\smallskip}
         & NGC\,5905 & IC\,3599 & NGC\,5905 & IC\,3599 \\
       \noalign{\smallskip}
      \hline
      \hline
       \noalign{\smallskip}
 H$\gamma$ 4340 & 0.20 & 0.26 &  &\\       
 ~[OIII] 4363 & 0.21 & 0.32 &  & \\       
 HeII 4686 & - & 0.37 &  & 870 \\
 H$\beta$ 4861 & 1 & 1 & 270 & 480 \\
 ~[OIII] 5007 & 0.49 & 3.2 & 300 & 260 \\
 ~[FeVII] 5721 & - & 0.10 &  & 440\\
 ~[FeVII] 6087 & - & 0.23 &  & 360\\ 
 ~[OI] 6300 & 0.30 & 0.18 &  & 430\\
 H$\alpha$ 6563 & 6.4 & 5.2 & 280 & 480 \\
 ~[NII] 6584 & 2.7 & 1.6& 290 & 430\\
 ~[SII] 6716 & 0.86 & 0.3 & 280 & 210 \\
 ~[SII] 6731 & 0.66 & $^{*}$ & 280 & \\ 
      \noalign{\smallskip}
      \hline
      \noalign{\smallskip}
  \end{tabular}

\noindent{\small $^{*}$ overlaps with atmospheric O$_{2}$\,$\lambda$6880 absorption.} 
   \end{table}

\paragraph{Line profiles.}
The emission line widths were determined by fitting Gaussians 
to the lines. The instrumental width was derived from night-sky lines 
which yielded 2.9$\pm{0.3}$\AA. 
The lines are very narrow, of order 270--300 km/s FWHM (Tab. 2; not corrected
for instrumental resolution).
In particular, we find no hints for an underlying broad emission
component that would indicate permanent low-level Seyfert activity.   
All properties of the emission line spectrum
are fully compatible with spectra observed in HII galaxies. 

\begin{figure} 
\psfig{file=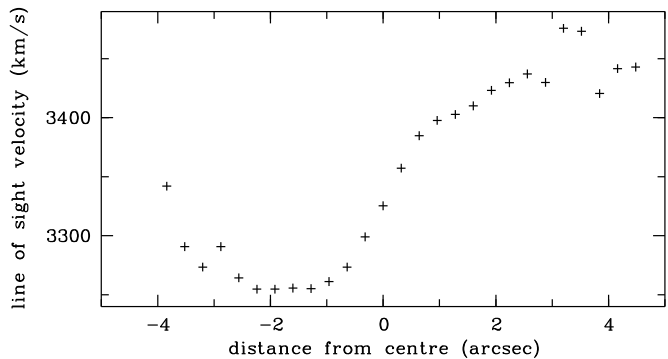,width=8.7cm}
\caption[vrot]{Line of sight velocities of NGC\,5905 measured in E-W
  direction with a slit width of 1\arcsec.}
\label{vrot} 
\end{figure}

\begin{figure*}[t]
\psfig{file=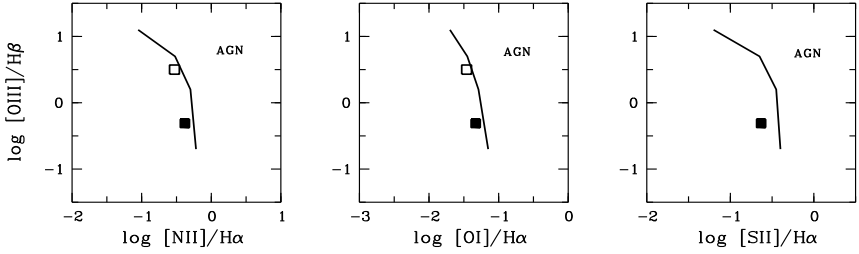,width=11.5cm,height=3.4cm}
\hfill
\begin{minipage}[]{0.32\hsize}\vspace*{-4.7cm}    
\hfill
 \caption[diagn]{Locations of NGC\,5905 (filled square) and IC\,3599 
(open square) in the standard diagnostic diagrams,
derived from single-component Gaussian fits to our emission line spectra.
}
\label{diagn}
\end{minipage}
      \vspace{-0.2cm}
\end{figure*}

We find clear signs of a rotation curve in the spectra of NGC\,5905.
The real (un-broadened) width of the lines is therefore even smaller
than given above.
To derive the rotation curve, we extracted 
one dimensional spectra of NGC\,5905.
The center shows a decreased H$\alpha$\ emission line
strength which can be interpreted with absorbing gas along the line of sight.
The E-W orientation of the slit is approximately
perpendicular to the orientation of the bar.
To extract line of sight velocities along the slit we measured the position of
the emission lines for each CCD row near the galaxy center.
A Gaussian was fitted to each emission line to determine its wavelength.
The distance between
the rows ($0\farcs 32$) is  clearly smaller than the seeing of $1\farcs 2$.
Therefore fine structure and a higher gradient in the rotation curve is
expected for spectra with
better spatial resolution. The highest rotation velocity is reached  2\arcsec\
outside the center which
corresponds to 0.72\,kpc. 
The velocity curve is shown in Fig. \ref{vrot} and further discussed in Sect. 7.3.
%
%

\subsubsection{IC\,3599}

\begin{figure} 
\psfig{file=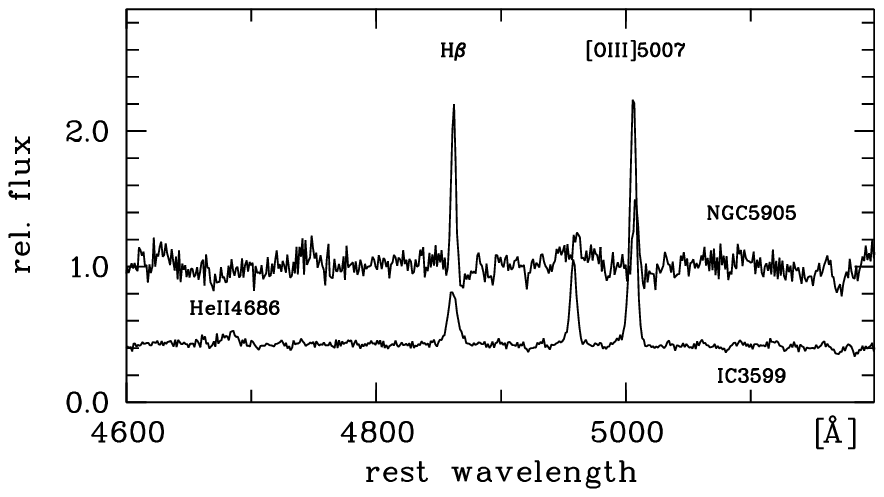,width=8.7cm}
\caption[]{\label{varspekblau} Spectrum of IC\,3599 (bottom) 
between 4600 and 5200\,\AA\ rest wavelength. The one of 
NGC\,5905 (top) is shown for comparison. 
IC\,3599 clearly shows
broader emission lines.}     
\vspace*{0.3cm}
\psfig{file=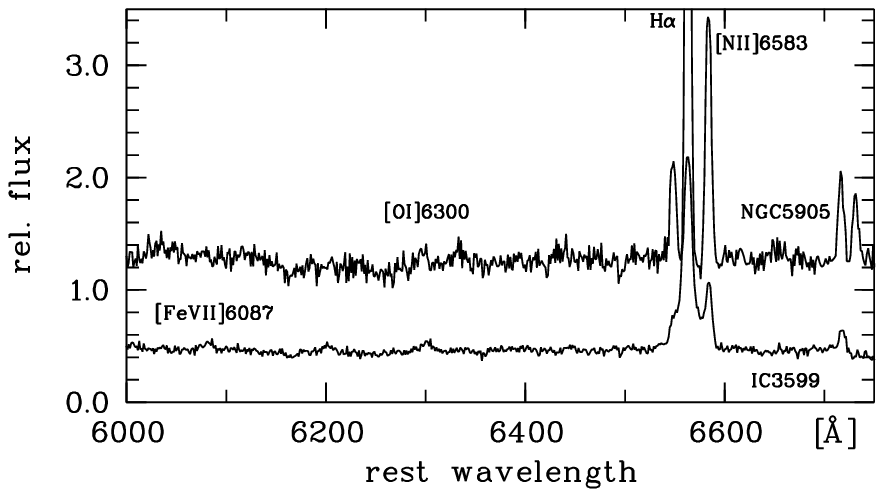,width=8.7cm}
\caption[]{\label{varspekrot} Spectra of IC\,3599 (bottom) and NGC\,5905
  (top) between 6000 and 6750\,\AA\ rest wavelength.}
\end{figure}

\paragraph{Emission line strengths.}
The optical spectrum of IC\,3599 differs from that of NGC\,5905 in several
respects (Figs. \ref{varspekblau}, \ref{varspekrot}). 
It shows stronger [OIII] emission and some 
high-ionization lines, like HeII$\lambda$4686 and [FeVII]$\lambda$6078.  
These 
have not changed in strength in comparison to the 1995 post-outburst spectrum in G\&95,  
indicating these lines may be permanent.  
We do not find signs for emission from FeII. 

\paragraph{Line profiles.}
The instrumental width was determined to be 4.5$\pm{0.5}$\AA. Line widths
given below and in Tab. \ref{obsres} are corrected for this value (4.5\AA).  
The Balmer lines have larger widths than the forbidden lines of [OIII] and [SII],
but the differences in line widths and the width of H$\beta$
(480\,km/s) would still be consistent with a Seyfert 2.
However, the complex of H$\alpha$ and the two [NII]
lines is significantly  better described if we assume {\em two} Gaussian components
for H$\alpha$ with FWHMs of $\rm 400\,km\,s^{-1}$ and $\rm 1090\,km\,s^{-1}$,
 both of similar intensity.{\footnote {Fitting {\em Gaussians} to emission
lines to assess the presence of a broad component is standard practice which also we
followed; `real' line profiles may be of more complex
shape, though.}}   

These spectral results obtained by our new spectra with higher spectral 
resolution
suggest an intermediate-type Seyfert classification 
for IC\,3599, namely a Seyfert\,1.9 in the scheme of Osterbrock (1981). 

\section {Photoionization models for the optical outburst spectrum of IC\,3599} 

\subsection {Model assumptions} 

In order to model the high-ionization emission line component 
in the optical outburst spectrum of
IC\,3599 (taken $\sim$0.5 yr after the observed X-ray outburst; BPF95), 
thereby deriving the physical properties of the illuminated gas,
we have calculated a sequence of photoionization models, again using the code
{\em Cloudy}. We assume constant density gas ionized by a pointlike
central continuum source and employ the mean Seyfert spectrum described in Sect. 3.1.2 
except for modifications in the EUV and X-ray range to account for the observed X-ray
spectral and variability properties.  

Emission line ratios were extracted from the spectrum of BPF95.
A large grid of models was calculated, varying all relevant unknown parameters
(i.e., the EUV--X continuum shape, the density $n$, the column density $N$, 
and the distance from the nucleus $r$; abundances were fixed
to the solar value) with the aim of 
matching the observed line-ratios during outburst
within a factor of 2 (note, e.g., uncertainties
in collisional strengths of coronal lines). We also carefully checked 
(and it turned out to be important) that no
other, non-detected, lines were overpredicted by the models (like 
[OIII], [OI]){\footnote {The high-density gas discussed below could 
contribute to [OIII],  
but clearly not to [SII] (much lower critical density). Therefore, we have 
checked that  the observed ratio of [OIII]/[SII] is constant from outburst to quiescence,
i.e. there is no strong contribution to [OIII] in outburst.}}.  

\subsection{Results} 
We find that the high observed X-ray flux in form of a strong `hot' soft excess
and a large gas column density are required to
produce the observed wide range in ionization states (like O$^0$ and Fe$^{13+}$). 
For the emission in the Fe lines there are always two solutions, one with
low density, the other with high density. The {\em high} density solution 
is needed to successfully
account for the strength of OI\,$\lambda$8446 (which is underpredicted otherwise)
and to avoid the overprediction of other low-ionization forbidden 
lines.       

The best match  
is obtained
for 
clouds of column density $N_{\rm H} \simeq 10^{23}$ cm$^{-2}$,
a gas density $n_{\rm H} \simeq 10^{9}$ cm$^{-3}$, and 
a cloud distance from the nucleus of $r \simeq 0.1$ pc.
Observed and modeled emission lines are compared in Tab. 3.
The derived parameters
are typical for BLR or CLR clouds{\footnote {for
some recent BLR models for Seyfert galaxies see, e.g., 
Radovich \& Rafanelli (1994), Baldwin et al. (1995); for coronal line region (CLR) models,
e.g.,    
Ferguson et al. (1997), Binette et al. (1997).}} 
suggesting that IC\,3599 permanently harbours a BLR
that is usually too weak to be easily detected (or is mainly hidden) 
but was illuminated by the X-ray outburst.

   \begin{table}[h]
\caption{Comparison of observed (BPF95)
and modeled
outburst emission lines ($I/I_{\rm H\beta_{\rm out}}$, `out' for outburst)
of IC\,3599.
   }
\begin{center}
     \label{fitres}
      \begin{tabular}{rcc}
      \noalign{\smallskip}
      \noalign{\smallskip}
      \hline
      \noalign{\smallskip}
        & \multicolumn{2}{c}{$I/I_{\rm H\beta_{\rm out}}$}  \\
      \noalign{\smallskip}
        Line~~~ & observed & modeled  \\
       \noalign{\smallskip}
      \hline
      \hline
       \noalign{\smallskip}
 HeII 4686 & 0.36 & 0.38 \\
 H$\beta$ 4861 & 1.0 & 1.0 \\ 
 HeI 5876  & 0.14 & 0.12 \\
 ~[OIII] 5007 & $<$ 0.2~~~ & 0.02 \\
 ~[FeX] 6375  & 0.37 & 0.45 \\
 ~[FeXI] 7892 & 0.23 & 0.36 \\
 ~[FeXIV] 5303 & 0.17 & 0.11 \\
 ~OI 8446 & 0.23 & 0.10 \\
      \noalign{\smallskip}
      \hline
      \noalign{\smallskip}
  \end{tabular}
\end{center}
   \end{table}


In traveling through the inner regions of the galaxy, the outburst emission 
acts as a probe of this region. It will be interesting to see 
when the emission reaches the distance, where in active galaxies the bulk of the NLR is located.

\section {Discussion: NGC\,5905 outburst scenarios}

The X-ray emission of normal spirals is too low to account for the 
maximum X-ray luminosity observed and drastic variability is not expected
(see also paper I).
Below, we scrutinize scenarios that may explain the observations. 

\subsection {Supernova}

X-rays are first emitted during the SN outburst itself.
Although energetically feasible, the timescale is too short 
(seconds to hours, instead of days) to account for the observed
X-ray outburst in NGC\,5905. 
On a longer time-scale, further radiation originates in a shock
produced by expansion of the SN ejecta into the nearby circumstellar 
medium.
The X-ray luminosity of the few detected SNe is much weaker
and the spectra are harder than that of NGC 5905, though.{\footnote { 
Only a few SNe (nearly all of type II) have been observed in X-rays so far 
(reviewed by Schlegel 1995). 
The maximal observed luminosities 
reached in the soft X-ray band are (Schlegel 1995, assuming $H_{\rm o}$ = 75)
2.7 10$^{39}$ erg/s (SN 1980K),
3.2 10$^{40}$ erg/s (SN 1986J),
9.5 10$^{39}$ erg/s (SN 1978K),
2.8 10$^{40}$ erg/s (SN 1993J, 5\,d after explosion), 
$\approx$ 10$^{41}$ erg/s (SN 1988Z; Fabian \& Terlevich 1996),  
and 1 10$^{39}$ erg/s (SN 1979C; Immler et al. 1998). 
The temperatures assuming a thermal spectrum 
range from $\sim$0.5 keV to $>$ 7 keV (Canizares et al. 1992, 
Bregman \& Pildis 1992, Schlegel 1995, Zimmermann et al. 1994b).
}}

\subsection {SN in dense medium}

The possibility and consequences of the existence of SNe in {\em dense} 
molecular gas was studied by Shull (1980) and
Wheeler et al. (1980). The idea was revived by Terlevich (1992) in the context of a starburst
interpretation of AGN.    

The basic scenario is the following:  
The matter ejected by the SN explosion expands into a region of high interstellar gas density.
The interaction  
causes a shocked region of hot gas enclosed by two shock waves, on the outside of the outgoing
shock, and on the inside of the inward reverse shock. The outward shock encounters
dense circumstellar material.
After sweeping up a small amount of gas 
these remnants become strongly radiative ($t = t_{\rm rad}$) and deposit most of their kinetic energy
in very short time-scales, thus reaching very high luminosities. Because of the large shock
velocities 
most of the energy is radiated in the EUV and X-ray region.
For expansion into {\em dense} material, 
all evolutionary phases 
are substantially
speeded up as compared to the `standard' case of $n \approx 10^0$ cm$^{-3}$.  

Since in this scenario high luminosities can be reached, and the evolutionary time
is considerably shortened (only of the order of years to few tens of years), a SN in dense
medium may be an explanation for the observed X-ray outburst in NGC\,5905.
In the following we first give some estimates using
the analytical treatment presented in Shull (1980) and
Wheeler et al. (1980).

Assume SN ejecta of initial total energy $E = 10^{51}E_{51}$ erg, expanding into
a medium of density $n = 10^4n_4$ cm$^{-3}$.
Most of the energy is emitted during/shortly after the onset of the radiative phase, 
which starts after a time
\begin {equation}
t = E_{51}^{0.22} n_4^{-0.56} (83\, \rm y)~~. 
\end {equation}
The evolution of temperature, radius and luminosity of the shock is then given by
\begin {equation}
T = E_{51}^{0.14} n_4^{0.27} (t/t_{\rm rad})^{-10/7} (2.7\, 10^7\, \rm K)~~,
\end {equation}
\begin {equation}
R= E_{51}^{0.29} n_4^{-0.43} (t/t_{\rm rad})^{2/7} (0.29\, \rm {pc})~~,
\end {equation}
\begin {equation}
L= E_{51}^{0.78} n_4^{0.56} (t/t_{\rm rad})^{-11/7} (9.8\, 10^{39}\, \rm {erg/s})~~,
\end {equation}
assuming line cooling to dominate with a cooling 
function of $\Lambda \propto T^{-0.6}$,
as expected from the X-ray fit.

Since it is unclear in which evolutionary state NGC\,5905 has been observed
we make the following assumptions:\\
(1) We assume the observed luminosity to be the peak luminosity
(and \NH = $N_{\rm H}^{\rm gal}$), dominantly emitted
in the X-ray regime.  These assumptions go in the direction of requiring
less extreme conditions.
We then check, if, and under which conditions, the observed high 
luminosity, and the quick decline in $L$
can be reproduced within this scenario. 
An energy of $E = 10^{51}$ erg is typically deposited by the explosion.
At $t = t_{\rm rad}$, and for a measured $L_{\rm x} \approx 3 \times 10^{42}$ erg/s,
the above equations can be solved for the density, yielding
$n_4 \simeq L_{\rm x}/(2.5 \times 10^6 L_{\odot}) \Rightarrow n = 3 \times 10^6 \rm {cm}^{-3}$.
A decrease in luminosity
within 5 months  
(1.45 yr) 
of a factor of $\sim 14$ (20) is then predicted, which compares to the observed drop
of a factor of $>$5 (100).   
Whereas these timescales roughly agree, 
the expected temperature is $T \simeq 10^8$ K, suggesting a {\em much harder}
X-ray spectrum than is observed. \\
(2) In a second step, we drop the assumption that $L$ was observed at its peak value,
and use the temperature derived from the X-ray Raymond-Smith (rs) fit as an estimate
of $T$, with $T \approx 7 \times 10^5$ K, $1 \times 10^6$ K for
the rs model with free or fixed $N_{\rm H}$, respectively.
In both cases the estimated densities
are unrealistically large, $n_{\rm H} > 10^{10}$ cm$^{-3}$. 

In conclusion, extreme conditions 
would be required and the predicted $T$ disagree.
Additionally, fine-tuning in the column density of the surrounding medium
would be needed in order to prevent the SNR from being completely self-absorbed.
Thus, we consider this scenario very unlikely.

\subsection {Accretion disk instabilities}
If a massive BH exists in NGC\,5905, it usually has to accrete with a very low
accretion rate or radiate with a low efficiency, or be completely dormant.
One solution in the framework of accretion disks that has recently received a lot
of attention, are advection dominated disks  
in which most energy is advected with the accretion flow instead of radiated away
(e.g., Abramowicz et al. 1988, Narayan \& Yi 1994; review by Narayan et al. 1998). 
This may be the dominant accretion mode in low-luminosity
objects (like in NGC\,5905 in `quiescence'). 

An instability in the accretion disk may then provide an explanation for the observed X-ray outburst.
Thermally unstable slim accretion disks were studied by, e.g., Honma et al. (1991),
who find the disk to exhibit burst-like oscillations for the case of the standard $\alpha$
viscosity description and for certain values of accretion rate. 
The high-luminosity state is characterized by a high mass accretion
rate in the inner disk region, lasting for a time of 
\begin{equation}
t_{\rm burst} \simeq 1.9\, \alpha_{0.1}^{-0.64} (M/M_{\odot})^{1.36} ~{\rm sec}
\end{equation}
where $\alpha$ = 0.1$\alpha_{0.1}$ is the standard viscosity 
parameter (e.g., Frank et al. 1985)
and $M$ is the mass of the central black hole.  
(For assumptions of this analytic result see Honma et al. 1991; for numerical 
results for a wider range of parameters cf. 
their Fig. 11). 
If the duration of the outburst is less than 5 months (i.e. 
the survey variability is taken as a real
rise in $L$), the mass of the central black hole in NGC\,5905 should 
be less than $\sim 10^5 M_{\odot}$.

These burst-like oscillations are found by Honma et al. only for certain
values of the initial accretion
rate.
A detailed quantitative comparison with the observed outburst in NGC\,5905 is difficult,
since the behaviour of the disk is quite model dependent, and further detailed modelling
would be needed.

\subsection {Warm absorber}

In this scenario, NGC\,5905 would host an active nucleus that is usually hidden
by a large column density of cold absorbing gas.
The Seyfert nucleus is intrinsically variable, and becomes visible
(in X-rays) during its flux high-states by ionizing the ambient medium that
becomes a {\em warm} absorber and transparent to soft X-rays.
This could naturally explain the observed variability as well as the steepness
of the outburst spectrum.

  \begin{figure}[t]    
\psfig{file=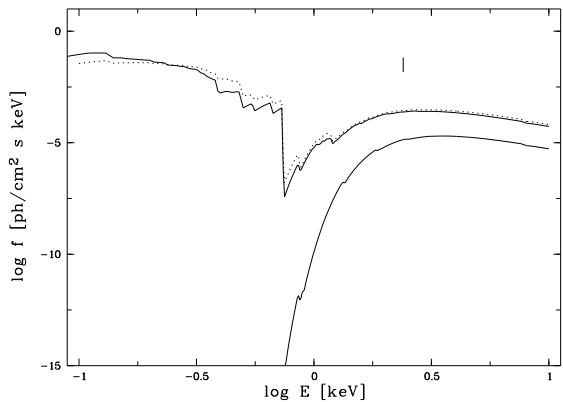,width=8.7cm}
 \caption[abs]{
Best-fit warm absorber model of the `outburst' spectrum of NGC\,5905
(upper solid curve), compared to the predicted spectrum resulting from a decrease in the
intrinsic luminosity by a factor of 10 for constant column density $N_{\rm w}$
(lower solid curve). The vertical bar marks
the end of the \ros energy band. The predicted countrate for the low-$L$ model is
below the detection limit in the pointed \ros observation.
The dotted line represents the dusty model for comparison
(for silicate only, 0.05 depleted).
}
\label{abs}
\end{figure}

We found that the high-state spectrum can be well described by a (dust-free)
warm absorber.
A source-intrinsic change in luminosity by a factor of less than 10 is needed
to change the absorption to complete within the \ros band (Fig. \ref{abs}),
veiling the view of the nucleus in soft X-rays.
Variability of such order is not unusual in low-luminosity AGN.

Since there is no evidence for Seyfert activity in the {\em optical} spectrum,
the nucleus must be hidden in this spectral region.
Assuming dust to be mixed with the total column density of the
warm absorber would result in an optical extinction of $A_{\rm v} \approx 34^{\rm m}$
(for a Galactic gas/dust ratio)
and would hide the Seyfert nucleus completely.
However, this dust strongly modifies the X-ray spectrum and such a model does
no longer fit the X-ray observation.
As we saw in Sect. 3.1.2, only fine-tuning in the dust properties enabled a
successful spectral fit in terms of a dusty warm absorber.
Alternatively, a steep underlying intrinsic spectrum could be assumed
to compensate for the flattening effect of dust. This is not an elegant
solution, though, since part of the original aim was to {\em explain}
the steep observed spectrum.

To summarize, a warm absorbed hidden Seyfert nucleus only explains the
observation when strongly fine-tuning the dust-properties.

\subsection {Tidal disruption of star}

Depending on its trajectory, a star gets tidally disrupted after passing a
certain distance to the black hole, the tidal radius, and 
the debris is accreted by the hole. 
This produces a flare of electromagnetic radiation. 
The peak luminosity should be a substantial fraction of the Eddington luminosity.
The timescale after which the bulk of the material is swallowed is estimated
to be of the order of months, after that, 
the luminosity declines as $\propto t^{-5/3}$ (Rees 1990).
Detailed numerical simulations of the   
disruption process (e.g., Laguna et al. 1993, Diener et al. 1997)
and consequences for the surroundings (e.g., Rauch \& Ingalls 1998)  
have been carried out in the last few years.
Explicit predictions of the emitted spectrum and luminosity 
are still rare. 
In particular, it is uncertain whether the emission peaks in the 
UV spectral region (e.g., Loeb \& Ulmer 1997, Ulmer et al. 1998) or in the
EUV -- soft\,X-rays (e.g., Rees 1988, Sembay \& West 1993).

If the X-ray outburst of NGC\,5905 is interpreted as a tidal disruption event,
the black hole mass can be roughly estimated via
the Eddington luminosity which is given by  
\begin{equation}
 L_{\rm edd} \simeq 1.3 \times 10^{38} M/M_{\odot} ~{\rm erg/s}~~.   
\end{equation}
In case of NGC\,5905, a BH mass of $\approx 10^{5}$ M$_{\odot}$ would be sufficient to
produce the observed $L_{\rm x}$ (assuming \NH = $N_{\rm H}^{\rm gal}$ and 
$L$ to be observed near maximum; this provides a {\em lower} limit on the BH mass).
In order to reach this luminosity,
$L_{\rm acc} = \eta {\dot M} c^2$, a mass supply of about 1/2000 M$_{\odot}$/year
is necessary, assuming $\eta$=0.1.
Rees predicts the event to last of the order of several months, consistent
with the observations of NGC\,5905 (cf. Fig. \ref{outlight}).

Again, a more detailed analysis has to await availability of more sophisticated
models. In particular, the flares cannot be standardised. Observations 
would depend on many parameters, like the type of disrupted star, the impact
parameter, the spin of the black hole, effects of relativistic precession,
 and the radiative transfer is complicated
by effects of viscosity and shocks (e.g., Rees 1994).  

\subsection {Further possibilities}

{\em Gravitational lensing event.}
High magnification factors can be achieved by lensing if the
lense lies nearly exactly on the line between observer and source
(see Schneider, Ehlers \& Falco 1992).{\footnote {
Ostriker \& Vietri (e.g., 1985) and others discussed the possibility 
to explain the strong and rapid variability of BL Lac objects 
via gravitational lensing. As a good candidate for lensing, Nottale (1986) 
presented the 
BL Lac object 0846+51W1 and successfully modeled the optical lightcurve 
which varied by about 4$^{\rm m}$ within one month.} 
If the X-ray variability of NGC\,5905 was caused by gravitational lensing
one would expect 
the same magnification factor in the optical as in X-rays
(assuming emission regions of similar size);
and the spectral component observed in the X-ray low-state has to
be of different origin then, since the spectral shape changed.
This contradicts the optical observations. Gravitational lensing therefore 
seems to be very unlikely. 

\vspace{0.3cm}
\noindent{\em X-ray afterglow of a GRB.}
For completeness, we finally note that within the time interval of  
the observed X-ray outburst 
no gamma-ray burst (GRB) has been reported{\footnote{The
possible connection of GRBs with nearby galaxies was discussed by, e.g., 
Larson et al. (1996), but see Schaefer et al. (1997). Recently, evidence 
has accumulated that GRBs are extragalactic and may 
originate in starburst regions of galaxies (e.g., Halpern et al. 1998).
For the suggestion that GRBs may be extreme tidal disruption events,
see Carter (1992).}}
(the 2 detected GRBs of July 8, 1990, have different positions;
Castro-Tirado 1994 on the basis of Granat/WATCH data).
We made the same check for IC\,3599. No simultaneous GRB is reported.
The one nearest the time of the X-ray outburst (Dec. 9, 1990)
of Dec. 19, 1990, has a different position.

\subsection {Summary of `successful' outburst scenarios} 

All those models that fullfil the available observational constraints
plus the order-of-magnitude estimates presented above involve the
presence of a SMBH in the center of the galaxy.
A particularly tight constraint is the huge observed outburst luminosity $L_{\rm x}$
of {\em at least} several 10$^{42}$ erg/s.     
The actual peak luminosity may even be much higher since (i) we may not have 
observed exactly at maximum light, (ii) the X-ray spectrum may extent into the EUV,
and (iii) there may be excess absorption (we usually assumed $N_{\rm H} = N_{\rm Gal}$).  
We favour the scenario of tidal disruption of a star because it can account
for the high outburst luminosity,  
seems to require least fine-tuning, and the long-term X-ray lightcurve
shows a continuous fading of the source over the whole measured time
interval (Fig. \ref{outlight}).
We re-caution, though, that many details
of the actual process of tidal disruption of a star and the emitted spectrum 
are still rather unclear.  

  \begin{figure*}[ht]
\psfig{file=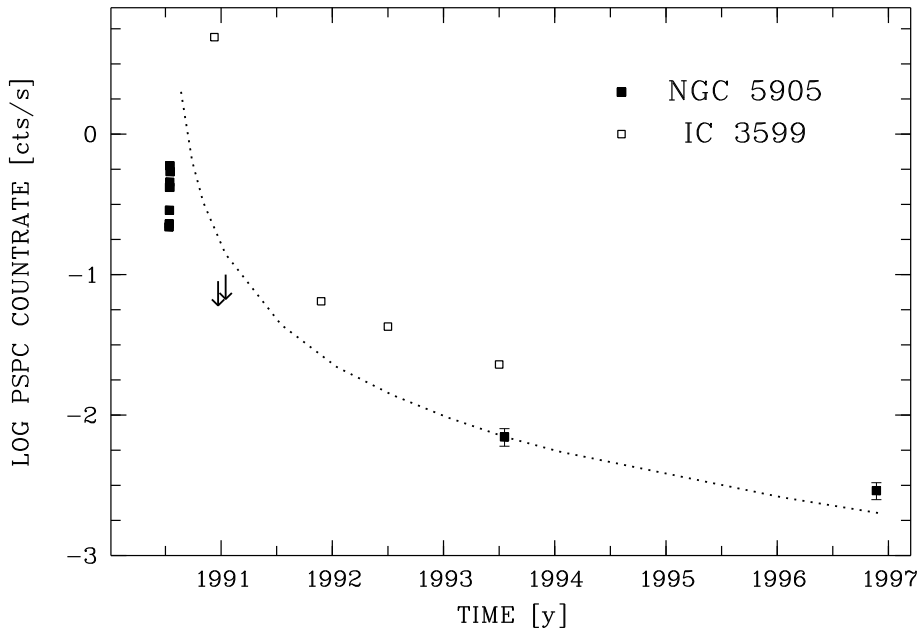,width=12.2cm}
\hfill
\begin{minipage}[]{0.26\hsize}\vspace*{-7.4cm}    
\hfill
 \caption[outlight]{Long-term X-ray lightcurve of NGC\,5905 (filled squares, and arrows
 for upper limits) 
 compared to IC\,3599 (open squares; these countrates  
 were taken from G\&95).
 The dotted curve is shown for illustrative purposes only. 
 It follows the relation $CR = 0.044 (t - t_{\rm o})^{-5/3}$ with 
 $t_{\rm o}$=1990.54.  
   }
\label{outlight}
\end{minipage}
\end{figure*}

\section {Discussion: Properties of NGC\,5905 in quiescence} 

\subsection{Classification}

Since the optical spectra obtained in 1997 (6.7 yr after the X-ray outburst)
do not show evidence for an high-excitation emission line component
we assume these spectra to represent the properties
of NGC\,5905 in quiescence. 
We have carefully searched for signs
of permanent low-level Seyfert activity.
This question is important in the context of AGN formation 
and evolution (e.g., Rees 1989), i.e., 
have we found evidence (via the X-ray outburst) for a SMBH in a galaxy that
otherwise  
appears perfectly non-active to an observer ?{\footnote{Evidence
for a population of non-active galaxies of unusually high X-ray luminosity
has occasionally been reported.   
However, optical follow-up
observations revealed signs of Seyfert activity in each spectrum
(e.g., Moran et al. 1994, 1996, Wisotzki \& Bade 1997).}} 

We find the emission lines to be very narrow, and lowly excited, as typical
for HII-type galaxies; we do not detect high-ionization lines like [FeVII],
and we do not find any signs for the presence of a broad component in the
Balmer lines. 
(Such a statement, of course, always reflects the current instrumental
sensitivity limits.)
The same results hold for a pre-outburst optical spectrum given in
Ho et al. (1995).

\subsection{HII properties} 

Second, we determine some of the properties of the low-ionization (HII-like)
emission line component. 

{\sl Extinction from Balmer lines.} 
The observed Balmer decrement is above its recombination value, indicative
of reddening by dust intrinsic to the emission line gas or within the host
galaxy along the l.o.s. Assuming all dust to be located in a screen 
outside the emission line gas and an intrinsic H$\alpha$/H$\beta$ ratio of
2.7 (Brocklehurst 1971)
yields an extinction of $E_{\rm B-V}$ = 0.75$^{\rm m}$  
for a Galactic extinction law as in Osterbrock (1989).

{\sl Density via sulphur ratio.} 
 The observed intensity ratio [SII]$\lambda$6716/6731  
implies a density of
the emission-line gas of $n \simeq 10^2$ cm$^{-3}$ (Osterbrock 1989), a value 
typical for HII regions.   

{\sl Temperature via oxygen ratio.} 
The oxygen intensity ratio [OIII]$\lambda$5007/4363 is unusually high. 
Taken at face value, it implies a temperature of $T > 20\,000$ K;  
but see Storchi-Bergmann et al. (1996) who find that 
the strength of [OIII]$\lambda$4363 
tends to be overestimated due to a contribution from the galactic stellar
spectrum.

\subsection{Rotation curve} 

The velocity curve we obtained (Fig. \ref{vrot}) allows a first rough estimate
of the mass enclosed within the central region.
The inclination angle of the galaxy is only known with a large
uncertainty because NGC\,5905 appears nearly face-on. We assume $ i =
30\degr$ and apply the formula given in Rubin et al. (1997), 
\begin{equation}
M = 2.33\,10^{5}\,v_{\rm rot}^{2} R ~(M_{\sun})~~,
\end{equation}
with $v_{\rm rot}$ in 
km/s and R in kpc. 
These values lead to $\rm 9\,10^{9}\,M_{\sun}$
within 0.7\,kpc. 
HST spectra would allow to improve this limit and thus the constraints
on a central dark mass.  

\section {Discussion: Comparison with IC\,3599}

Some confusion exists in the literature about the classification 
of IC\,3599 on the basis of optical spectra.
The optical (post-outburst) spectrum was classified as
Seyfert\,2 by G\&95, as starburst by Bade et al. (1995) and Grupe (1996),
whereas BPF95 classified the (outburst) spectrum
as NLSy1-like.  
We originally selected IC\,3599
as a second candidate for a non-active galaxy 
that showed an X-ray outburst. 
Below, we comment on the classification on the basis 
of emission line strengths and profiles derived from our high-resolution spectra.  

The emission-line ratios fall on the dividing line between Seyfert and HII galaxies 
in the diagnostic diagrams of Veilleux \& Osterbrock
(1987; Fig. \ref{diagn}). 
However, [OII] is weaker than [OIII] (cf. Fig. 8 of Bade et al. 1995), as typical 
for Seyferts, whereas HII galaxies (and also LINERs) usually show [OII]/[OIII] $>$ 1. 
HeII is observed in some LINERs (even HII galaxies; e.g.
Osterbrock 1989), 
but the value observed in IC\,3599 is
very high even for Seyfert 2s (cf. the [OIII]-HeII correlation in 
Komossa \& Schulz 1997).  
In outburst, the spectrum may share some similarities with a NLSy1 (BPF95), but we note
that the small width of the Balmer lines would lead to the expectation
of strong Fe\,II emission (given the general trends seen in NLSy1s), 
which is {\em not} observed.   

Concerning line profiles, in our new spectra with higher resolution 
than previously obtained  we find a second broad component in H$\alpha$
that argues for a Seyfert\,1.9 classification of IC\,3599
(at the epoch of observation, and within the `standard' classification
schemes). 
We also note that there is similarity to the
`composite' classification of several objects in Moran et al. (1996).

Alternatively, the high-ionization lines as well as broad H$\alpha$ 
may be relics from the outburst. However, the high-ionization lines have not changed 
in strength as compared to the post-outburst optical spectra of G\&95
which may indicate they are permanent. Further monitoring 
of these emission lines may be worthwhile.  
But even neglecting 
line profiles and the high-ionization component, 
the low-ionization component (particular the weakness of [OII]) still
suggests a Seyfert classification.   

Our photoionization models for the optical outburst emission lines
further corroborate evidence that IC\,3599 permanently possesses
a BLR or CLR.  

In X-rays, the lightcurves of both galaxies are strikingly similar (Fig. \ref{outlight}).
A potential problem for the tidal disruption scenario for IC\,3599 may be
the following, though. If IC\,3599 is active and permanently possesses an accretion
disk, the disk's presence may hinder the disruption process.

\section {Discussion: Search for, and comparison with, other strongly variable objects}

Besides NGC\,5905 and IC\,3599, non-recurrent X-ray outbursts{\footnote {
we use the term `outburst' to indicate variability stronger than a factor
$\sim$50 }} 
have been reported for E1615+061 (Piro et al. 1988) and WPVS007 (Grupe et al. 1995b)
and repeated variability of large amplitude was found in 
IRAS\,13224 (Otani et al. 1996, Boller et al. 1997).  
All these latter galaxies are Seyfert galaxies  
as shown by their optical
emission line spectra. 
Different mechanisms to account for the X-ray variability have been favoured
for the active galaxies:
A variable soft excess for E1615+061 (Piro et al. 1988, see also Piro et al. 1997)
and WPVS007 (Grupe et al. 1995b), relativistic effects in the accretion disk for IRAS\,13224
(Brandt et al. 1997, Boller et al. 1997).    

It is also interesting to point out the following: high-amplitude variability
seems to 
go hand in hand with the presence of a warm absorber.  
We performed a search for further cases of strong X-ray variability (Komossa 1997) 
using the sample of nearby galaxies of Ho et al. (1995) and \ros survey
and archived pointed observations.   
We do not find another object with a factor $\sim$100 amplitude, 
but several sources are discovered to be strongly variable (with maximal factors ranging 
between $\sim$10 and 20 in the mean countrates); 
NGC\,3227 (Komossa \& Fink 1997c), NGC\,4051 (Komossa \& Fink 1995) and
NGC\,3516. 
All of them are known to harbour warm absorbers  
(NGC\,4051, e.g., Mihara et al. 1994, Komossa \& Fink 1997b; NGC\,3227, 
Ptak et al. 1994, Komossa \& Fink 1997c; 
NGC\,3516, Mathur et al. 1997). Two further examples are 
NGC\,3786 (Komossa \& Fink 1997d) and NGC\,3628 (Dahlem et al. 1995; who 
detect a drop of a factor 20 in \ros flux and suggest several different
explanations including the possibility of a warm absorber in this object).    
All of these are known to be Sy galaxies (except NGC\,3628) and it is unclear at present,
whether there is a direct link between the presence of a warm absorber 
and strong variability; we do not favour the warm absorber interpretation
for the outburst of NGC\,5905.

In any case, it is important to note that large-amplitude variability
seems to appear in {\em all} types of AGN, not only in NLSy1s.

\section{Summary and conclusions}

We have presented follow-up optical and X-ray observations of NGC\,5905 and IC\,3599
and discussed outburst scenarios. 

Our high-resolution optical spectra do not reveal any signs of
Seyfert activity   
in NGC\,5905 and classify the galaxy as HII type, as judged by emission line widths and strengths.
In contrast, IC\,3599 shows some high-ionization
emission lines and evidence for a broad component in
H$\alpha$, suggesting a Sy 1.9 classification (at the epoch of observation), 
although at present we cannot exclude
that the broad component is still a relic from the outburst.  
The optical spectrum of NGC\,5905 shows the pattern of a rotation curve,
which allowed a first estimate of the mass enclosed within 0.7 kpc:
$M \approx 10^{10}$ M$_{\odot}$.  

Photoionization modelling of the optical outburst spectrum of IC\,3599 
implies a gas density of $n \simeq 10^9$ cm$^{-3}$ and a
distance from the nucleus $r \simeq$ 0.1 pc of the emission line gas illuminated during
the outburst and thus corroborates evidence that the galaxy permanently
possesses a BLR or CLR.  

Optical photometry of NGC\,5905 quasi-simultaneous to the X-ray outburst
and on longer terms does not reveal optical variability.
Constraints on the long-term development of optical brightness
after the X-ray outburst  
could be improved by using the excellent spatial resolution of the HST.  

New HRI observations of NGC\,5905, performed 6.3 yrs after the outburst,
indicate a further drop in X-ray flux by a factor $\sim$2 as compared
to the last PSPC observation.  

Several outburst scenarios have been examined. 
Tight observational constraints are the giant amplitude of variability,
at least a factor $\sim$100, and the huge outburst luminosity of at least
several 10$^{42}$ erg/s. 
All successful models involve the presence of a SMBH. 
The most likely scenario to explain the X-ray outburst in NGC\,5905
seems to be tidal disruption of a star by a central SMBH; a scenario
proposed by Rees (1988) as a tracer of SMBHs in nearby galaxies.
We caution, though, that many theoretical details of this process are still rather
unclear.

Given the absence of any optical signs of 
Seyfert activity, at present this makes NGC\,5905 the {\em only} non-active galaxy
among the X-ray outbursting ones.

The X-ray outburst in this HII galaxy then lends further support to the scenario that
{\em all} galaxies passed through an active phase (instead of just a few), leaving
dormant SMBHs in their centers.

\begin{acknowledgements}
St.K. acknowledges support from the Verbundforschung under grant No. 50\,OR\,93065. 
It is a pleasure to thank C. LaDous and G. Richter at {\sl Sternwarte
Sonneberg} for their kind
hospitality and G. Richter for valuable help in
the assessment of the photographic plates. 
We thank Gary Ferland for providing {\em{Cloudy}}; Jochen Greiner, Niel Brandt and Andrew Ulmer
for discussions and comments; and Niel Brandt for providing an enlargened copy 
of Fig. 4 of BPF95.
The \ros project is supported by the German Bundes\-mini\-ste\-rium
f\"ur Bildung und Wissenschaft (BMBW/DLR) and the Max-Planck-Society.

\end{acknowledgements}

\end{document}